\documentclass[11pt]{article} % font size
\usepackage[T2A]{fontenc}
\usepackage[utf8]{inputenc}
\usepackage[english]{babel}
\usepackage[round]{natbib} 
\usepackage{aas_macros}% этот макрос позволяет правильно отобразить ссылки на статьи из NASA ADS в списке литературы

\usepackage{adjustbox}
\usepackage{titling}
\usepackage{fancyhdr}

\usepackage[a4paper,margin=1in]{geometry} % page setup, margins
\usepackage{authblk} % author affiliations
\usepackage[colorlinks=true, linkcolor=black, citecolor=black, urlcolor=blue, filecolor=blue,breaklinks=true]{hyperref} % Hyperlinks, needed for biblatex, other option: allcolors=blue
\usepackage{setspace} % for linespacing
\usepackage{lineno} % line numbering (change with \linespacing below)
\usepackage{csquotes} % Recommended for biblatex, must load after lineno or will get a warning
\usepackage{ragged2e} % for left aligning
\usepackage[capitalise, noabbrev, nameinlink]{cleveref}
\crefdefaultlabelformat{#2\textbf{#1}#3} % To bold figure refs
\Crefname{figure}{\textbf{Figure}}{\textbf{Figures}} % To bold figure refs
\Crefname{section}{\textbf{Section}}{\textbf{Sections}} % To bold section refs
\Crefname{table}{\textbf{Table}}{\textbf{Tables}} % To bold table refs
\usepackage{graphicx}
\graphicspath{{./main_figs/}{./supplement/suppl_figs/}} % paths for figs
\DeclareGraphicsExtensions{.pdf,.jpeg,.JPG,.png,.PNG, .eps, .tiff}
\usepackage{subcaption} % for subcaptions (panels), and add parentheses
\DeclareCaptionLabelFormat{bold}{\textbf{(#2)}} % bold subpanel letter in caption
\usepackage{pgffor} % for function to make figures with subpanels
\usepackage{alphalph} % for function to make figures with subpanels
\usepackage[labelfont=bf, textfont=bf, singlelinecheck=off, textfont=footnotesize]{caption} % boldness of captions
\usepackage{booktabs}
\usepackage{multirow}
\usepackage{rotating}
\makeatletter
\renewcommand{\AB@affillist}{}
\renewcommand{\AB@authlist}{}
\makeatother

\title{Dynamical investigation of the multiple star ADS 9173 AB}

\author[1]{O.V.~Kiyaeva\footnote{e-mail:kiyaeva@list.ru}}
\author[1]{I.S.~Izmailov}
\author[1]{N.V.~Narizhnaya}
\author[1]{L.G.~Romanenko}
\affil[1]{The Central Astronomical Observatory of the Russian Academy of Sciences at Pulkovo}

\date{Jule 2023} 
\begin{document}

%\RaggedRight % If you want left-aligned (ie ragged right text) rather than justified
\setstretch{1.15} 
\maketitle

\thispagestyle{fancy}
\begin{abstract}
Star ADS 9173=WDS 14135+5147=Hip 69483 is a complex system. The B component has a spectroscopic companion, whose orbit with a period of 4.9 years has been known since 1986. The Gaia telescope has detected a distant faint pair over $100"$ away from the bright AB pair. In our article, we study the movement in a bright pair based on long-term observations with the 26-inch refractor of the Pulkovo Observatory. The AB pair orbit with a period of 6306 years was calculated using the apparent motion parameters (AMP) method. The astrometric orbit of the component B was determined on the basis of the residuals of the  homogeneous CCD observations up to 2023 with the 26-inch refractor. It is in agreement with the spectroscopic one. The remaining secondary residuals show a wave with a period of approximately 20 years, the reasons for which are discussed.
\end{abstract}

\section*{Introduction}
When studying the process of formation of stars and exoplanets, the further dynamic evolution of stellar groups and systems of exoplanets, it is important to pay attention to double and multiple systems. Observations in the optical to centimeter range indicate that multiple systems of two or more bodies are the norm throughout all stages of stellar evolution. It is widely accepted that multiple systems arise from the collapse and fragmentation of cloud cores, despite the inhibitory effect of magnetic fields. This fact is noted in many works on these topics \citep{Reip2014}. The search for multiple systems is therefore an urgent observational problem. Especially when it comes to studying exoplanet systems \citep{Magr2019}. 

Observation of double and multiple stars is a traditional topic for the Pulkovo Observatory since the days of W.~Struve. Observations and a search for possible inner subsystems are currently ongoing with the 26-inch refractor \citep{Shakht2010, Grosh2006, Kiya2021}. This article continues a series of ADS 9173 studies \citep{Kiya2006, opast} based on photographic and CCD observations with a 26-inch refractor.

In the work \citep{Kiya2006} the preliminary orbit of the AB pair was determined for the first time by the AMP method on the basis of photographic observations in 1982--2004. Also an attempt was made to obtain astrometric orbit consistent with the spectroscopic orbit of the photocenter, using residuals. The spectroscopic orbit was derived from 70 years of radial velocity observations as early as 1986 \citep{Bakos}. From the spectroscopic orbit, 3 dynamic parameters were adopted: period $P$, eccentricity $e$ and the instant of the periastron passage $T$. The astrometric orbit allows you to determine the orientation of the orbit plane (the angle of inclination $i$ and ascending node longitude $\Omega$). Orbit matching is controlled by a common parameter independently obtained spectroscopically and astrometrically, the periastron longitude from the ascending node $\omega$. The accuracy of the photographic observations was not sufficient to obtain reliable agreement, although the presence of a satellite in positional observations was noted. In addition, the perturbation in declination with a period of more than 13 years were discovered.

In the work \citep{opast}, based on photographic and CCD observations of 2003--2012, a preliminary orbit of a possible satellite with a period of 20 years was determined. At the same time, the motion of the AB pair in the 1982--2012 segment turned out perpendicular to the plane of the sky in the direction $\rho$, since $\dot{\theta}=0.0002 \pm 0.0010 \approx 0^\circ$/year, which is not confirmed by 200-year observations. It was not possible to obtain a satisfactory orbit of the AB pair with such parameters of apparent motion.

At present, the Gaia telescope has discovered by the common parallax one more pair of weak stars ($m\approx{15^m}$) Cab of low mass (see the MSC catalog \citep{MSC2018}) at a distance of $114''$. In the Gaia DR3 catalog \citep{G2022} there are no parallax and proper motions for the B component due to low accuracy which is created by the inner subsystem, and there is also no radial velocity of the A component. Perhaps the reason for the lack of data is that the stars are bright, and there are problems with determining the radial velocities for stars of spectral type A (according to the WDS catalog \citep{WDS}: $m_A=4.57^m, Sp_A=A7V$, $m_B=6.81^m, Sp_B=F1V$).  

In this work, the orbit of the AB pair is determined from the entire series of observations. From the residuals relative to this orbit the astrometric orbit of component B is determined. It consistents with the spectroscopic orbit within the error limits. The question of the reality of the second satellite with a period of 20 years, which manifests itself in residual deviations, is discussed. 

\section{Observations}
\begin{figure}
\centering
    \includegraphics[width = 0.49\textwidth]{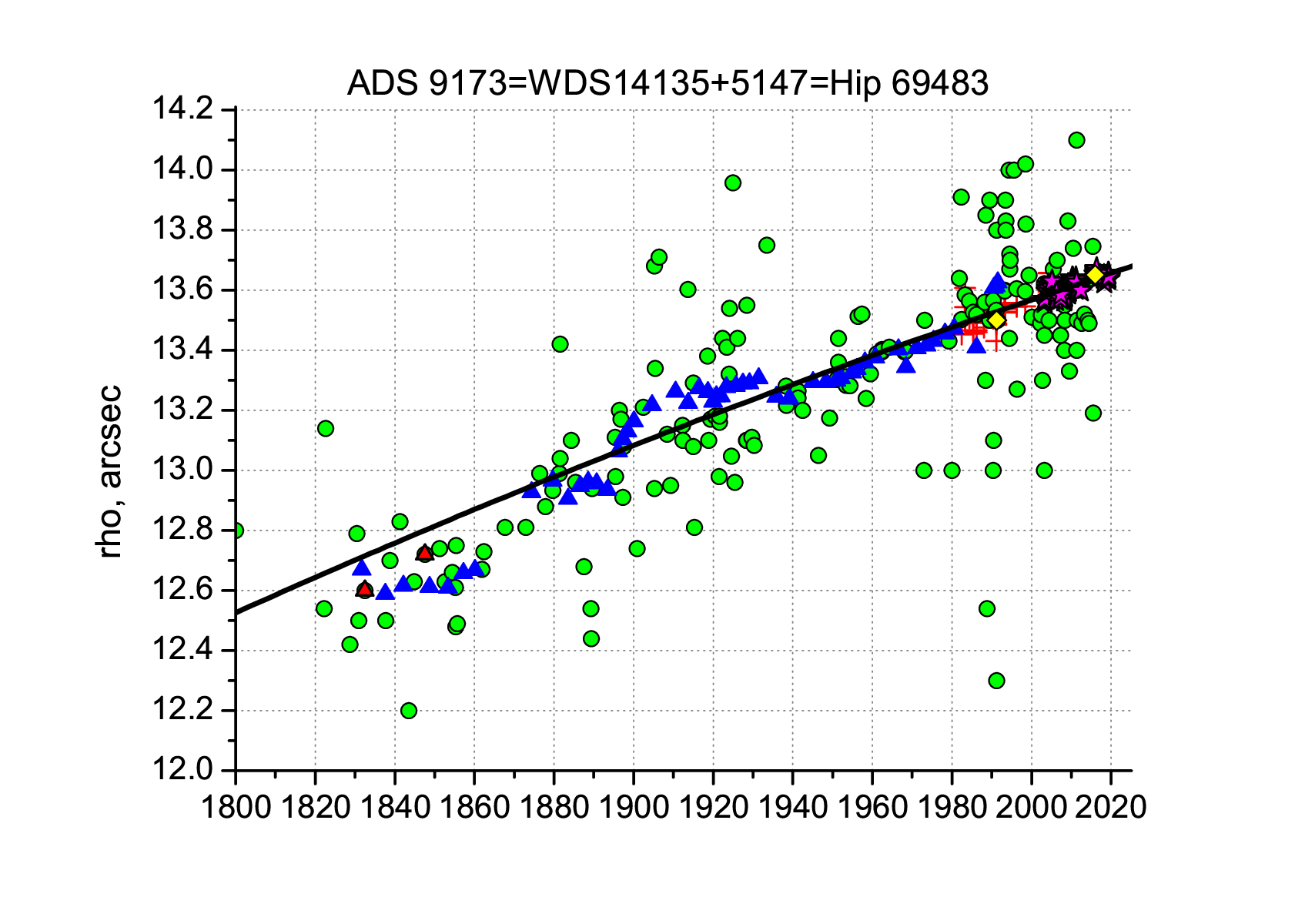}
    \includegraphics[width = 0.49\textwidth]{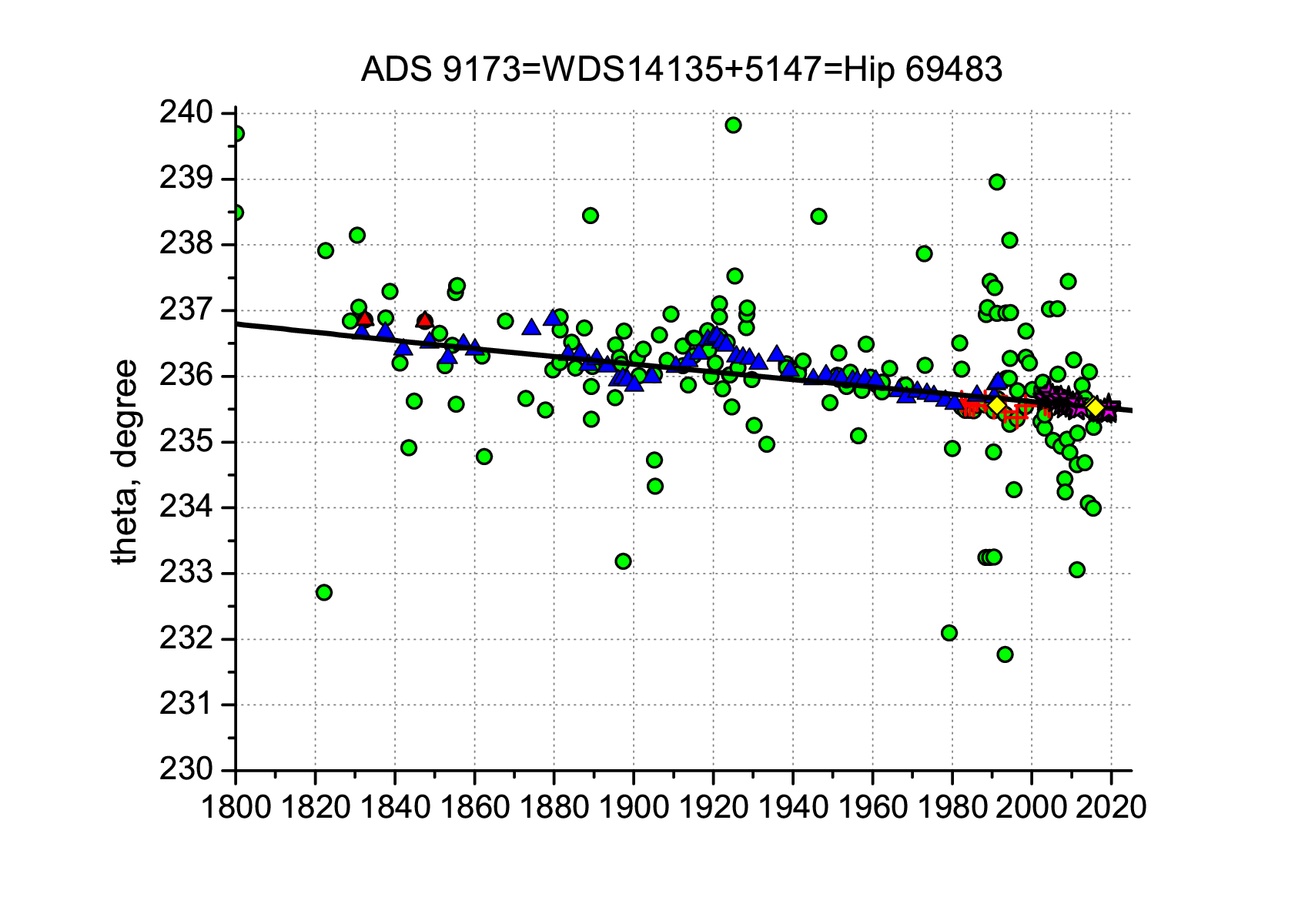}
\caption{The entire series of the ADS 9173 AB observations. Designations: green circles indicate observations from WDS \citep{WDS}, blue triangles are smoothed series, the red triangles are the observations of Wilhelm and Otto Struve, yellow diamonds are space observations of Hipparcos (from WDS), Gaia DR2 \citep{G2018} and Gaia DR3 \citep{G2022}, red crosses are Pulkovo photographic observations \citep{Cat2014}, magenta asterisks are CCDs (see \cite{Izm20}), the line is the ephemeris of 
the wide pair AB orbit.}
\label{fig-obsAB}
\end{figure}

The bright pair ADS 9173 AB was discovered by William Herschel in 1779, but W. Struve's 1832 observation (obtained by 7 separate positions) and O. Struve's 1847 observation (obtained by 11 separate positions) are the most reliable early observations. among those included in the WDS catalog.
There are only 192 observations in WDS (2016 version) for 1779--2015. During this time, the distance between the components has changed only by $1''$, and the position angle by $1^\circ$  (see Fig. \ref{fig-obsAB}). To reduce the large scatter caused by the influence of possible inner subsystems, we averaged all observations in a sliding window of 20 years. After 2000, we took into account only our homogeneous observations. Later, to determine the orbit of the outer pair, we used this series 1831--2010 (62 positions). 

%\newpage 

The star has been observed at the Pulkovo Observatory with a 26-inch refractor since 1982. There are 48 photographic observations from 1982--2004 measured with a scanner \citep{Cat2014}, and 71 CCD observations from 2003--2019 (see \cite{Izm20}). The last 13 CCD-observations for 2020--2023 are presented in this paper. A systematic difference between photographic and CCD-observations in $\rho$ was found: Photo-CCD=50 mas. This correction has been introduced in the set of photographic observations. Comparison of the combined series with the worldwide observations from the WDS did not reveal any systematic discrepancies.

\begin{figure}
\centering
    \includegraphics[width = 0.49\textwidth]{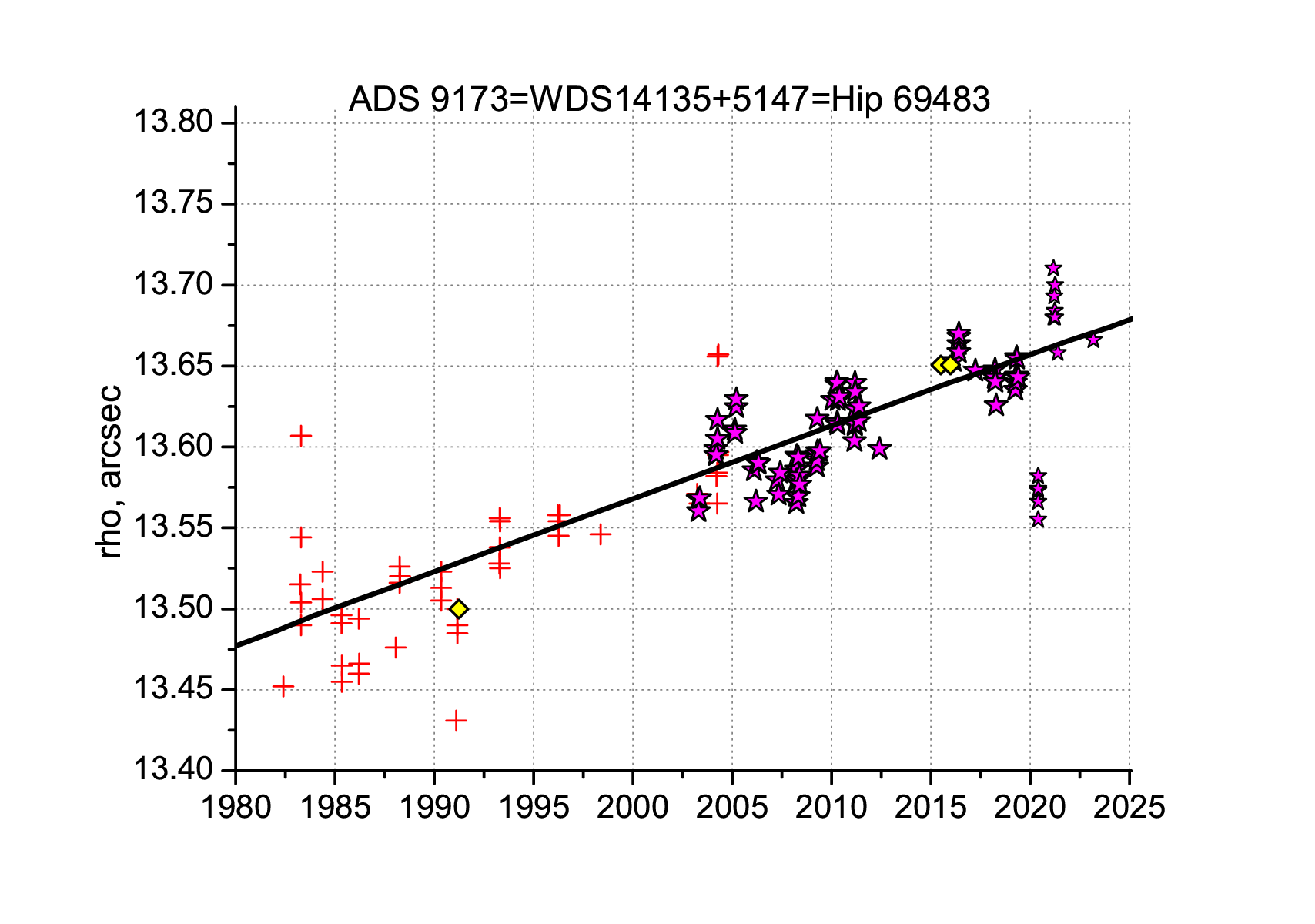}
    \includegraphics[width = 0.49\textwidth]{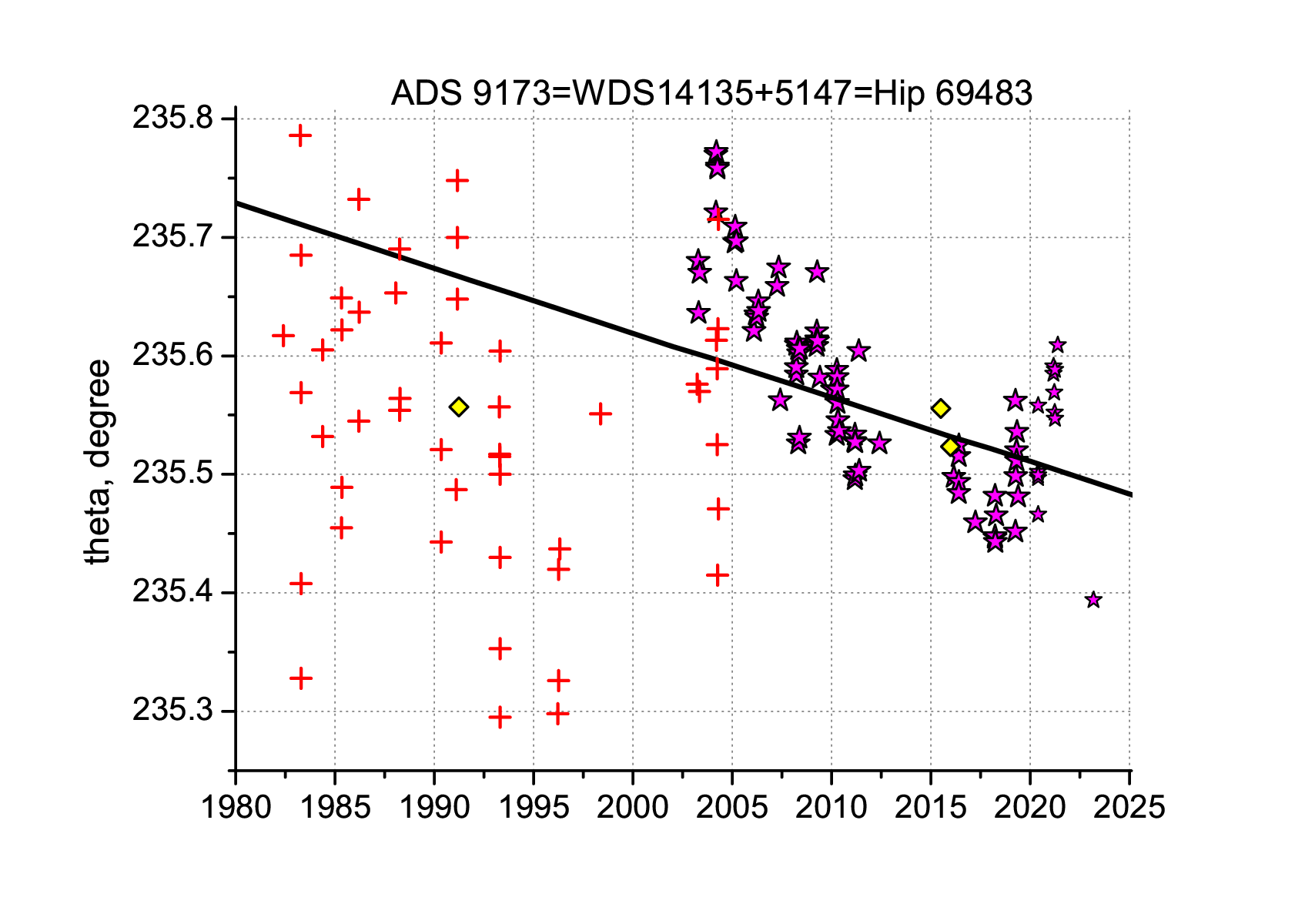}
\caption {A series of observations on the Pulkovo 26-inch refractor. The designations are the same as in Fig.~\ref{fig-obsAB}. Dependence $\theta(t)$ shows a disturbance with a period of about 20 years.} 
\label{fig-obspu}
\end{figure}

All observations are shown in Fig. \ref{fig-obsAB}. The Pulkovo observations are presented separately in Fig. \ref{fig-obspu}.

%\newpage 

On the $\rho(t)$ graph, the smoothed series demonstrates a clearly pronounced wave in the 1830--1940 segment, however, it doesn't show up any further. This example of observational selection shows that one wave is not enough to confirm the astrometric orbit obtained from residuals,   careful analysis of all available data is required.

\section{AMP-orbit of the A-B pair}
The AMP method \citep{KK1980} is designed to determine the initial orbits of wide visual binary stars with a long period of revolution in position and velocity at the one moment on the basis of the results of observations, obtained by various available methods.  These are the apparent motion parameters (AMP) at the instant $T_\circ$: the apparent distance between components ($\rho$), the position angle ($\theta$), the apparent relative motion ($\mu$), and the positional angle of direction of the relative motion ($\psi$), the radius of curvature ($\rho _c$).

In addition, the necessary data are the parallax $p_t$ (for the relationship between linear and angular quantities), the relative radial velocity of the components $\Delta{V_r}$ obtained from spectroscopic observations (to calculate the spatial velocity vector of the satellite B relative to the main star A) and an estimate of the components mass sum $\Sigma{M}$ according to the data on the physical properties of the stars. 

If it is possible to determine all five parameters, including the radius of curvature, then the distance between the components $r$ in astronomical units is calculated by the formula

\begin{equation}
\label{one}
 r^3=k^2\frac{\rho\rho_c}{\mu^2}|\sin (\theta -\psi)|
\end{equation}
where $k^2=4\pi^2\Sigma{M}$ is a dynamic constant if distance is measured in AU, time in years, mass in solar mass units. 

Then we get two position vectors that correspond to the position of the secondary component symmetrically with respect to the picture plane, and, consequently, two orbits, each of which is characterized by an angle $\beta$ between the spatial position of the satellite and its projection onto the picture plane.

\begin{equation}
\label{two} \beta=\pm\arccos{\frac{\rho}{rp_t}}
\end{equation}

Both orbits have the same semi-major axes and periods, but other parameters differ.

Most precisely, the AMP are obtained from homogeneous observations (basis), performed on the same telescope in order to eliminate instrumental systematic errors. All other observations are used to control and, if necessary, to refine the unknown initial data.

In our case, as can be seen from Fig. \ref{fig-obspu}, the most accurate is a series of CCD observations. However, the relative movement of the components is too slow, and it is impossible to get the parameters of the visible movement only on the basis of this homogeneous series, distorted by the influence of possible satellites. Therefore the AMPs are calculated for the moment 1950.0 over the entire smoothed series. The components mass sum we took from the MSC catalog. Unfortunately, it is impossible to determine the radius of curvature, and the relative radial velocity is determined uncertainly. 

In the work \citep{Bakos}, $V_{rB\gamma}=-21.5$ ~km/s was determined for component B and $V_{rA}=-23$~km/s was approximately estimated. In agreement with the remote observations, we determined the angle $\beta=-11^\circ$ and refined $\Delta{V_r}=+1.7$~km/s. The smoothed observations of 1831--1940 and 2000--2010 were used as distant observations. 

To select the best solution, we use an algorithm, proposed in the work \citep{Kiya1983}. In this  algorithm observations and ephemeris are compared not directly, but the Thiele-Innes elements A, B, F, G, calculated in two ways: by the geometric elements of the AMP orbit ($a, i, \omega$ and $\Omega$) and as a least-squares method solution of the system of linear equations, where, along with the observed coordinates, the dynamical elements of the AMP orbit ($P, T$ and $e$) are used. Ideally, these values should match. We are looking for the minimum of the comparison function S depending on the difference of the Thiele-Innes elements. 

\begin{equation}
\label{three}
S=(\Delta{A}^2+\Delta{B}^2+\Delta{F}^2+\Delta{G}^2)^{1/2}
\end{equation}

\begin{figure}
\centering
    \includegraphics[width = 0.49\textwidth]{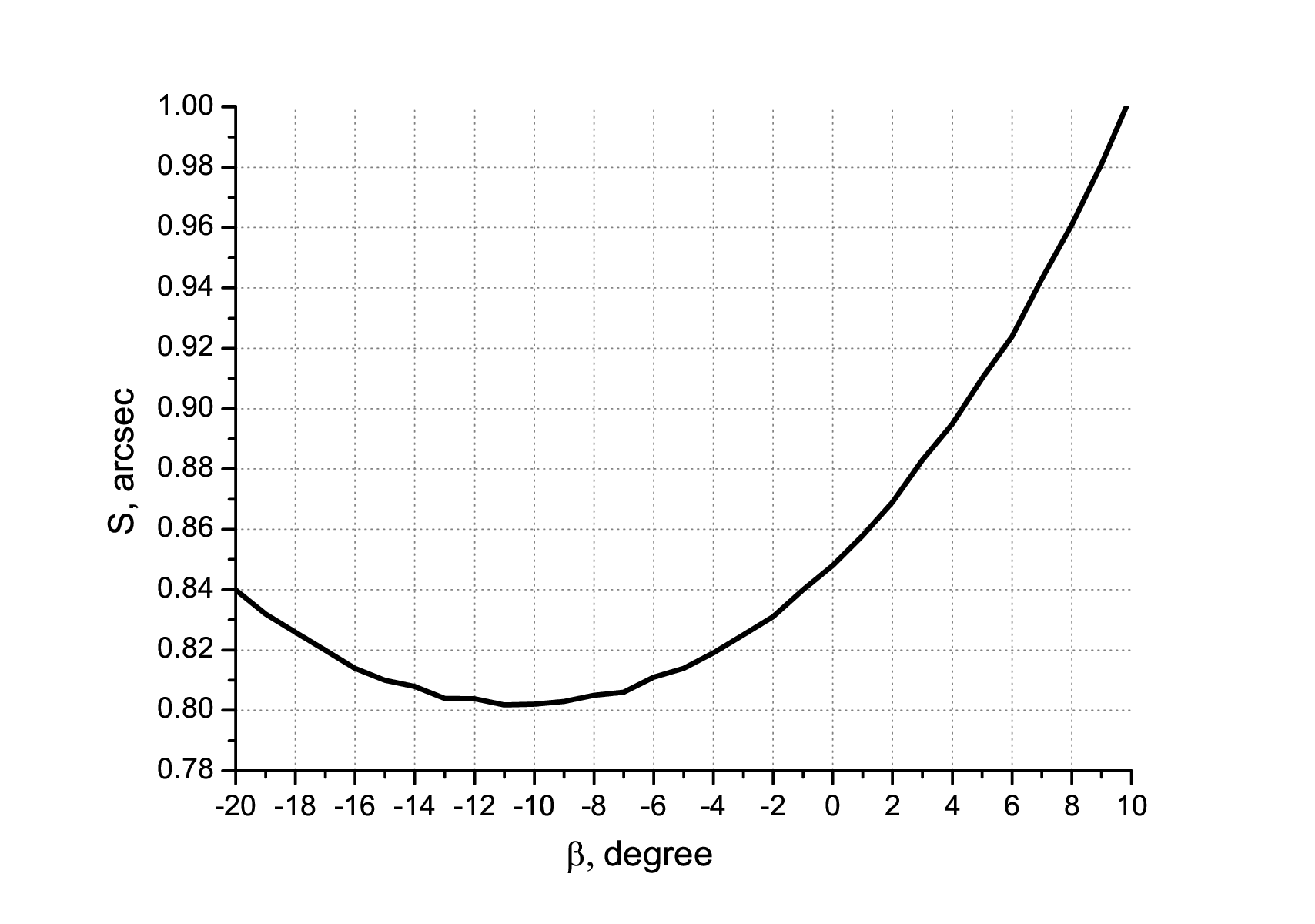}
    \includegraphics[width = 0.49\textwidth]{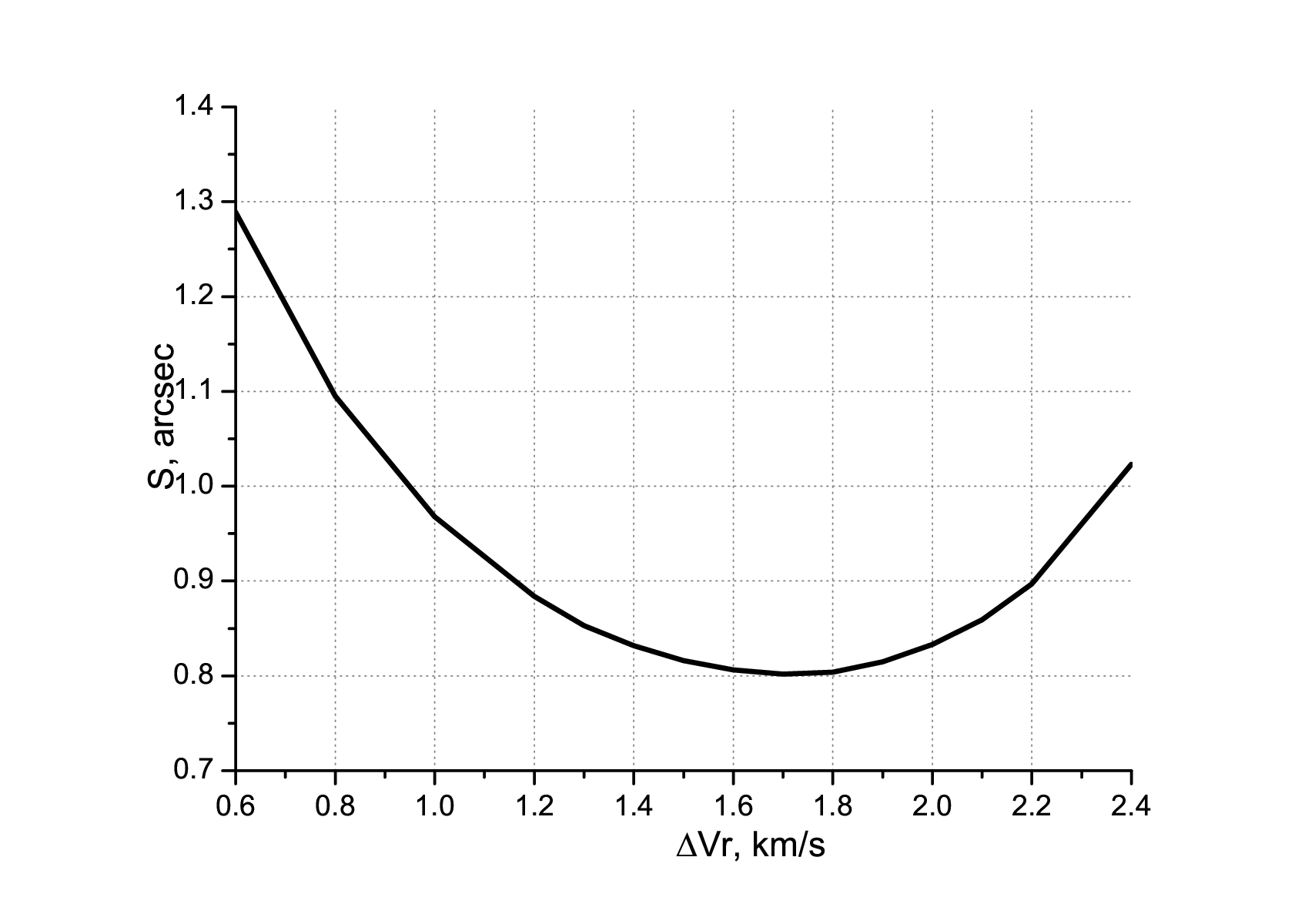}
\caption{ Dependences $S(\beta)$ and $S(\Delta{V_r})$ for ADS 9173 AB.} 
\label{fig-S}
\end{figure}

Dependences $S(\beta)$ and $S(\Delta{V_r})$ are shown in Fig. \ref{fig-S}. 

%\newpage 

According to this criterion, we confidently obtain the best solution corresponding to $\beta=-11^\circ \pm 7^\circ$. If we use the minimum dispersion criterion over the entire series of observations, the second solution, corresponding to $\beta = +11^\circ$, also passes well through the entire series. However, when comparing the Thiele-Innes elements, we take into account not only positional observations, but also the elements of the AMP-orbit $T$ and $e$, which are different for different solutions. Therefore, we believe that the only correct solution is the one corresponding to $\beta=-11^\circ$. 

\begin{figure}
    \centering
    \includegraphics[width = 0.49\textwidth]{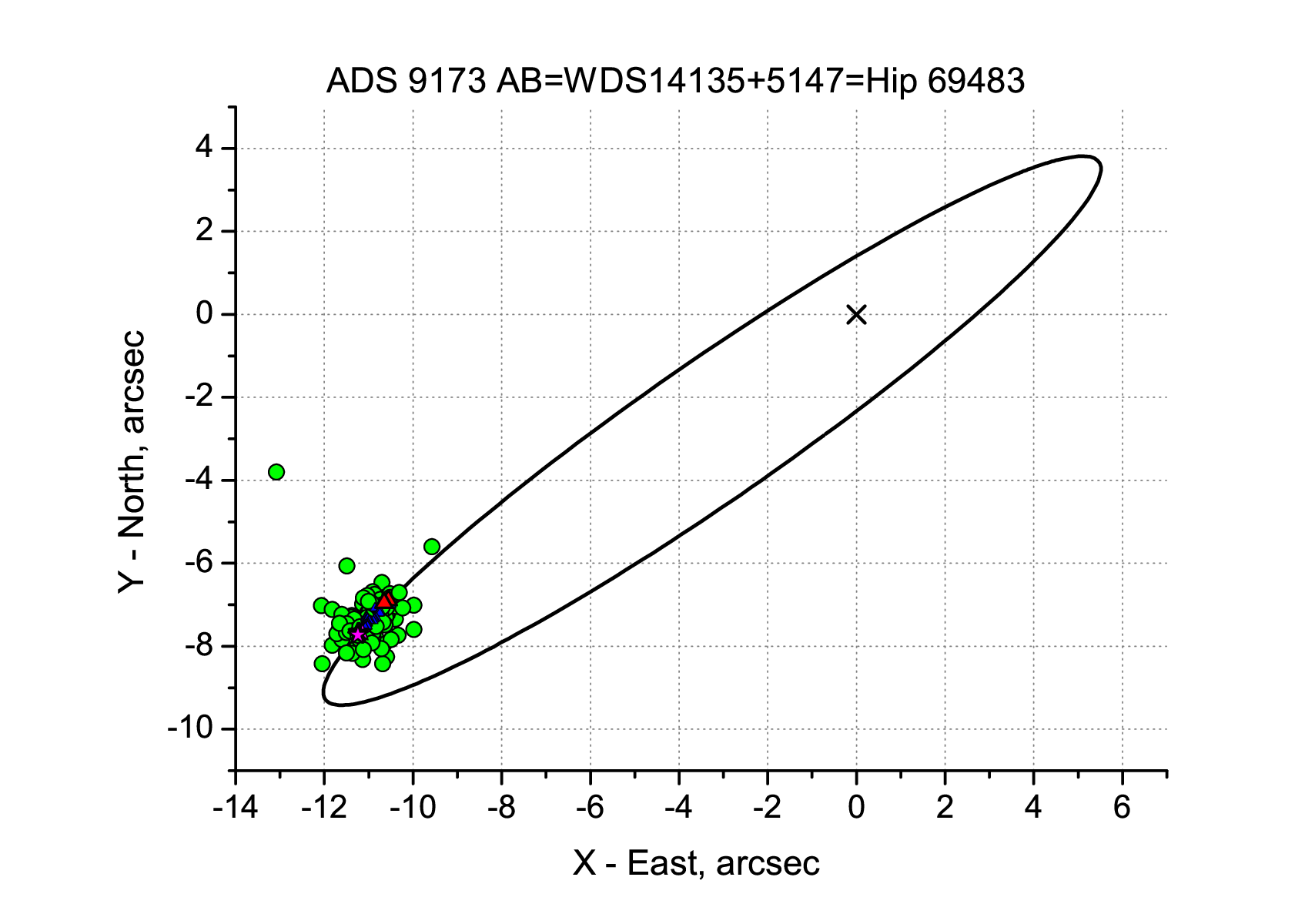}
    \caption{Orbit of the outer A-B pair in the picture plane. The oblique cross denotes the main component of A, the rest of the designations are the same as in Fig. \ref{fig-obsAB}.}
\label{fig-orbAB}
\end{figure}

%\newpage 
\begin{table}
\begin{center}
\caption{Initial data for determining the AMP orbit of the AB pair.}\label{tab-pvd}
\begin{adjustbox}{width=0.9\textwidth}
\begin{tabular}{cccccccc}
\hline
   $T_0$, yr & $rho, ''$ & $\theta, ^\circ$ & $\mu$, mas/yr& $\psi, ^\circ$ & $\Delta{V_r}$, km/s & $p_t$, mas    & $\Sigma{M}, M_\odot$ \\
\hline
 1950.0 & 13.335 &  235.897 &  5.029 & 220.70 & +1.7 &   20.153 &  4.2 \\
   & $\pm$.010 &  $\pm$ 0.023 & $\pm$.193 & $\pm$ 1.28 & $\pm$ 0.2 & $\pm$ 0.009 &      \\
\hline
\end{tabular}
\end{adjustbox}
\end{center}
\end{table}

\begin{table}
\begin{center}
\caption{Elements of the AMP orbit of the AB pair.} \label{tab-AB0}
\begin{adjustbox}{width=0.59\textwidth}
\begin{tabular}{r  rrrrrrr}
\hline
     & $a, ''$ & $P$, yr &    e & $\omega, ^\circ$  & $i, ^\circ$ & $\Omega, ^\circ$ & $T$, yr \\
\hline
     &  11.10  &  6306. &  .44 &  208.0 &   99.2 &  234.1 &  6515. \\
  +  &   1.56  &  1362. &  .10 &   16.5 &    1.1 &     .2 &   119. \\
  -  &    .96  &   800. &  .19 &   16.6 &    1.4 &    1.7 &   611. \\
\hline
\end{tabular}
\end{adjustbox}
 {\scriptsize 
 
 The second and third lines contain the uncertainties that define the range for each parameter.}
\end{center}
\end{table}

The initial data for obtaining the AMP-orbit are presented in Table \ref{tab-pvd}, orbit elements - in Table \ref{tab-AB0}.

Orbital element errors are determined by the total change of each element when changing all the initial parameters by the values of their errors. Since the resulting range of element values is not symmetrical with respect to the calculated parameter, we indicated the uncertainties in both directions. Fig. \ref{fig-orbAB} shows the orbit of the outer pair in the picture plane.

\section{Astrometric orbit of the inner Ba-Bb pair}

To determine the astrometric orbit of component B, we use only CCD observations 2003--2023. During this time, the pair made 4 revolutions. The observations and (O-C) in right ascension (dx1) and in declination (dy1) relative to the orbit of the AB pair are presented in Table \ref{tab-dxy}. The residuals dx1 and dy1 are the input data for calculating the astrometric Ba-Bb orbit.  Observation weights are determined by the $Err \rho$ error. 

\begin{table}[h!]
\begin{center}
\caption{CCD observations of the AB pair with a 26-inch refractor, and residuals.} \label{tab-dxy}
\begin{adjustbox}{width=0.99\textwidth}
\begin{tabular}{ccccccccc}
\hline
 t, yr   & $\rho, ''$ & $ Err \rho, ''$ & $\theta, ^\circ$ &
 $Err \theta, ^\circ$ & dx1, mas & dy1, mas & dx2, mas & dy2, mas \\
\hline
 2003.380 &  13.5683 &  .0034 &  235.6699 &  .0158 &    003.4 &   022.2 &   -2.8643  &  11.8928 \\
 2004.180 &  13.5949 &  .0056 &  235.7695 &  .0179 &   -029.5 &   029.6 &  -26.0504  &  22.9184 \\
 2004.188 &  13.5987 &  .0065 &  235.7208 &  .0258 &   -026.1 &   017.9 &  -22.5598  &  11.2674 \\
 2004.215 &  13.5951 &  .0066 &  235.7719 &  .0263 &   -029.9 &   030.1 &  -26.0534  &  23.6356 \\
 2004.259 &  13.6049 &  .0067 &  235.7595 &  .0161 &   -036.2 &   022.2 &  -31.8589  &  16.0161 \\
 2004.264 &  13.6166 &  .0095 &  235.7579 &  .0363 &   -045.7 &   015.4 &  -41.3033  &   9.2484 \\
 2005.111 &  13.5634 &  .0068 &  235.6285 &  .0230 &    018.1 &   023.1 &    29.1807  &  23.4550 \\
 2005.146 &  13.6108 &  .0057 &  235.6954 &  .0277 &   -029.9 &   009.6 &  -18.7407  &  10.2505 \\
 2005.149 &  13.6085 &  .0065 &  235.7087 &  .0216 &   -029.7 &   013.5 &  -18.5350  &  14.1763 \\
 2005.206 &  13.6245 &  .0083 &  235.6631 &  .0186 &   -036.7 &  -004.3 &  -25.4676  &  -3.1448 \\
 2005.217 &  13.6294 &  .0060 &  235.6965 &  .0202 &   -045.1 &  -000.5 &  -33.8633  &   0.7470 \\
 2006.102 &  13.5856 &  .0041 &  235.6209 &  .0128 &    003.7 &   012.6 &    -2.8461  &  14.7308 \\
 2006.195 &  13.5662 &  .0118 &  235.6348 &  .0418 &    018.1 &   026.6 &    8.2601  &  27.2435 \\
 2006.233 &  13.5908 &  .0060 &  235.6320 &  .0143 &   -001.7 &   012.3 &  -12.7180  &  12.3128 \\
 2006.329 &  13.5896 &  .0077 &  235.6457 &  .0211 &   -002.3 &   016.0 &  -15.8055  &  14.4550 \\
 2006.348 &  13.5901 &  .0053 &  235.6378 &  .0210 &   -001.5 &   014.2 &  -15.4203  &  12.3570 \\
 2007.268 &  13.5794 &  .0101 &  235.6586 &  .0168 &    007.2 &   027.6 &   -9.5144  &  17.4873 \\
 2007.344 &  13.5705 &  .0090 &  235.6742 &  .0302 &    012.8 &   036.0 &   -3.3668  &  25.6276 \\
 2007.421 &  13.5840 &  .0067 &  235.5621 &  .0177 &    016.9 &   006.7 &    1.3460  &  -3.8875 \\
 2008.229 &  13.5697 &  .0042 &  235.5838 &  .0188 &    028.1 &   021.9 &   21.1669  &  11.4492 \\
 2008.240 &  13.5859 &  .0048 &  235.5840 &  .0159 &    014.7 &   012.9 &    7.9001  &   2.4767 \\
 2008.243 &  13.5652 &  .0067 &  235.5900 &  .0144 &    031.1 &   025.8 &   24.3371  &  15.3844 \\
 2008.248 &  13.5794 &  .0055 &  235.6090 &  .0188 &    016.8 &   021.5 &   10.0979  &  11.0972 \\
 2008.259 &  13.5945 &  .0049 &  235.6109 &  .0174 &    004.1 &   013.3 &   -2.4685  &   2.9257 \\
 2008.327 &  13.5695 &  .0073 &  235.5259 &  .0229 &    036.3 &   011.1 &   30.5618  &   0.9155 \\
 2008.330 &  13.5934 &  .0106 &  235.6055 &  .0248 &    005.9 &   013.1 &    0.1978  &   2.9242 \\
 2008.357 &  13.5800 &  .0082 &  235.6017 &  .0204 &    017.6 &   020.1 &   12.2305  &  10.0064 \\
 2008.390 &  13.5821 &  .0047 &  235.5308 &  .0182 &    025.4 &   005.2 &   20.4359  &  -4.7888 \\
 2008.406 &  13.5181 &  .0094 &  235.4576 &  .0167 &    088.1 &   027.1 &   83.3330  &  17.1639 \\
 2008.409 &  13.5766 &  .0087 &  235.6056 &  .0172 &    020.0 &   022.9 &   15.2705  &  12.9740 \\
 2009.266 &  13.5954 &  .0051 &  235.6201 &  .0090 &    005.1 &   018.3 &   10.5631  &  12.7980 \\
 2009.269 &  13.5878 &  .0088 &  235.6081 &  .0181 &    013.0 &   020.2 &   18.4956  &  14.7187 \\
 2009.272 &  13.5918 &  .0049 &  235.6135 &  .0132 &    008.9 &   019.0 &   14.4268  &  13.5387 \\
 2009.283 &  13.6174 &  .0119 &  235.6701 &  .0391 &   -019.7 &   015.7 &  -14.0569  &  10.3139 \\
 2009.296 &  13.5974 &  .0064 &  235.6123 &  .0143 &    004.6 &   015.6 &   10.3806  &  10.3037 \\
 2009.414 &  13.5975 &  .0064 &  235.5814 &  .0207 &    008.9 &   010.0 &   15.8776  &   5.5402 \\
 2010.058 &  13.6291 &  .0064 &  235.5714 &  .0171 &   -013.9 &  -007.6 &   -2.7263  &  -6.8822 \\
 2010.279 &  13.6385 &  .0061 &  235.5878 &  .0160 &   -023.2 &  -008.8 &  -12.2681  &  -6.2679 \\
 2010.282 &  13.6399 &  .0084 &  235.5330 &  .0291 &   -017.0 &  -020.4 &   -6.0811  & -17.8450 \\
 2010.285 &  13.6398 &  .0119 &  235.5815 &  .0277 &   -023.4 &  -010.8 &  -12.4950  &  -8.2212 \\
 2010.299 &  13.6153 &  .0089 &  235.5714 &  .0338 &   -001.7 &   001.1 &    9.1377  &   3.7865 \\
 2010.318 &  13.6138 &  .0054 &  235.5598 &  .0159 &    001.0 &  -000.3 &    11.7341  &   2.5319 \\
 2010.323 &  13.6289 &  .0057 &  235.5452 &  .0146 &   -009.4 &  -011.6 &    1.3046  &  -8.7303 \\
 2010.413 &  13.6310 &  .0087 &  235.5362 &  .0331 &   -009.6 &  -014.2 &    .3985  &  -10.6868 \\
 2011.161 &  13.6035 &  .0113 &  235.5275 &  .0279 &    016.4 &   002.3 &    4.6962  &    1.9170 \\
 2011.178 &  13.6394 &  .0084 &  235.5331 &  .0169 &   -014.0 &  -016.9 &  -26.1629  & -17.5611 \\
 2011.183 &  13.6134 &  .0050 &  235.4954 &  .0156 &    012.6 &  -009.5 &   0.3057  & -10.2428 \\
 2011.186 &  13.6142 &  .0060 &  235.5270 &  .0243 &    007.7 &  -003.8 &  -4.6734  &  -4.5925 \\
 2011.205 &  13.6233 &  .0052 &  235.5266 &  .0120 &    000.3 &  -009.0 &  -12.5489  & -10.0986 \\

\hline
\end{tabular}
\end{adjustbox}
\end{center}
\end{table}

\begin{table}[h!]
\begin{center}
\begin{adjustbox}{width=0.99\textwidth}
\begin{tabular}{ccccccccc}
\hline t, yr   & $\rho, ''$ & $ Err \rho, ''$ & $\theta, ^\circ$ &  $Err \theta, ^\circ$ & dx1, mas & dy1, mas & dx2, mas & dy2, mas \\
\hline
 2011.208 &  13.6340 &  .0076 &  235.4989 &  .0213 &   -004.8 &  -020.4 &  -17.7232  & -21.5477 \\
 2011.385 &  13.6159 &  .0087 &  235.6037 &  .0312 &   -003.4 &   011.0 &  -19.5765  &   7.2251 \\
 2011.388 &  13.6249 &  .0046 &  235.5024 &  .0180 &    002.8 &  -014.0 &  -13.4149  & -17.8143 \\
 2012.423 &  13.5989 &  .0100 &  235.5258 &  .0311 &    024.0 &   009.0  &   9.2995  &  -1.7953\\
 2016.166 &  13.6530 &  .0060 &  235.4977 &  .0222 &   -005.9 &  -013.9  &  -20.0280  & -15.8974\\
 2016.409 &  13.6636 &  .0116 &  235.5154 &  .0342 &   -016.3 &  -015.5  &  -33.7144  & -20.7763\\
 2016.415 &  13.6681 &  .0071 &  235.4936 &  .0169 &   -017.1 &  -022.3  &  -34.5598  & -27.6451\\
 2016.417 &  13.6586 &  .0062 &  235.4841 &  .0141 &   -008.0 &  -018.8  &  -25.4743  & -24.1675\\
 2016.420 &  13.6699 &  .0053 &  235.5244 &  .0160 &   -022.7 &  -017.3  &  -40.1967  & -22.7023\\
 2017.242 &  13.6471 &  .0054 &  235.4595 &  .0130 &    007.2 &  -014.2  &   -8.2459  & -24.8191\\
 2018.236 &  13.6478 &  .0076 &  235.4464 &  .0204 &    011.3 &  -013.6  &   6.8038  & -23.4619\\
 2018.241 &  13.6426 &  .0126 &  235.4479 &  .0361 &    015.3 &  -010.3  &   10.8655  & -20.1446\\
 2018.247 &  13.6429 &  .0064 &  235.4819 &  .0224 &    010.5 &  -003.8  &    6.1395  & -13.6239\\
 2018.255 &  13.6403 &  .0046 &  235.4431 &  .0154 &    018.0 &  -010.0  &   13.7390  & -19.7956\\
 2018.299 &  13.6257 &  .0101 &  235.4655 &  .0171 &    027.1 &   002.8  &   23.3820  &  -6.8365\\
 2019.257 &  13.6352 &  .0094 &  235.5623 &  .0293 &    008.9 &   019.8  &   16.2159  &  15.5971\\
 2019.263 &  13.6398 &  .0079 &  235.4516 &  .0236 &    020.2 &  -004.5  &   27.5728  &  -8.6587\\
 2019.287 &  13.6436 &  .0065 &  235.4984 &  .0238 &    010.8 &   002.7  &   18.3991  &   -1.2797\\
 2019.325 &  13.6439 &  .0058 &  235.5202 &  .0156 &    007.6 &   006.9  &   15.5475  &   3.2062\\
 2019.328 &  13.6553 &  .0078 &  235.5116 &  .0234 &   -000.6 &  -001.3  &    7.3750  &  -4.9706\\
 2019.350 &  13.6544 &  .0053 &  235.5357 &  .0201 &   -003.1 &   004.1  &    5.0708  &   0.5969\\
 2019.410 &  13.6431 &  .0047 &  235.4814 &  .0199 &    013.8 &   000.0  &   22.4849  &   -3.0383\\
 2020.404 &  13.5664 &  .0108 &  235.4975 &  .0384 &    077.6 &   050.2  &   84.8754  &  54.6701\\
 2020.409 &  13.5818 &  .0127 &  235.4658 &  .0458 &    069.3 &   035.3  &   76.4714  &  39.7860\\
 2020.412 &  13.5731 &  .0137 &  235.5021 &  .0291 &    071.6 &   047.4  &   78.7096  &  51.8950\\
 2020.415 &  13.5740 &  .0081 &  235.4988 &  .0266 &    071.3 &   046.2  &   78.3446  &  50.7040\\
 2020.418 &  13.5554 &  .0096 &  235.5583 &  .0339 &    078.6 &   068.3  &   85.5816  &  72.8125\\
 2021.177 &  13.7096 &  .0073 &  235.5911 &  .0148 &   -050.8 &  -009.8  &  -66.7591  & -13.3571\\
 2021.185 &  13.6800 &  .0082 &  235.5847 &  .0145 &   -025.5 &   005.7  &  -41.5702  &   2.0328\\
 2021.231 &  13.6926 &  .0095 &  235.5688 &  .0267 &   -033.6 &  -004.4  &  -50.2398  & -8.6746\\
 2021.248 &  13.6842 &  .0099 &  235.5518 &  .0392 &   -024.3 &  -002.9  &  -41.1256  &  -7.3920\\
 2021.264 &  13.6802 &  .0097 &  235.5465 &  .0364 &   -020.2 &  -001.6  &  -37.1874  &  -6.2911\\
 2021.267 &  13.7003 &  .0106 &  235.5885 &  .0331 &   -042.5 &  -004.7  &  -59.5158  & -9.4272\\
 2021.395 &  13.6584 &  .0077 &  235.6091 &  .0289 &   -010.4 &   023.4  &  -28.3202  &  17.2279\\
 2023.205 &  13.6656 &  .0053 &  235.3938 &  .0173 &    017.9 &  -016.6  &   14.2069  & -26.2290\\
\hline
\end{tabular}
\end{adjustbox}
\end{center}
 {\scriptsize The residuals in right ascension (dx1) and declination (dy1) after taking into account the AMP orbit of the AB pair are used as initial data to determine the astrometric orbit of Ba-Bb. The  residuals dx2 and dy2 remain after taking into account also the astrometric Ba-Bb orbit.}
\end{table}

The photocenter orbit is defined in a standard way. We solve the system of equations: 

\begin{equation}
\label{four} 
dx_{phase}=dx_0+BX_{phase}+GY_{phase}
\end{equation}
\begin{equation}
dy_{phase}=dy_0+AX_{phase}+FY_{phase}
\end{equation}
where $dx_{phase}$ and $dy_{phase}$ are the original residuals depending on the phase with respect to the period, which is equal to the fractional part of $(t-T_0)/P$; $X_{phase}=cos(E_{phase})-e$, $Y_{phase}=\sqrt{1-e^2}sin(E_{phase})$ are the orbital coordinates, depending on the orbital elements $P, T$ and $e$, which in our case are known with high accuracy from the spectroscopic orbit \citep{Bakos}, $E$ is the true anomaly at time t. Note that the  instant of the periastron passage $T$ is recalculated taking into account the fact that from 1927 to 2010 the star made 17 revolutions. 

We determine the coordinates of the mass center $dx_0$ and $dy_0$ at the instant $T_0=2012.0$ and the elements Thiele-Innes A, B, F, G, which correspond to the elements of the orbit $(a,i, \omega, \Omega)$. The errors of these elements relative to the calculated solution were determined by the Monte Carlo method, the number of trials is 30, the initial residuals were distorted with an adopted dispersion of 15 mas, which corresponds to the accuracy of the CCD observations. These errors are slightly larger than those of the average solution, difference between model and average elements falls within this range.

The results are shown in Table \ref{tab-AB} and in Fig. \ref{fig-orbB}. 

\begin{figure}
    \centering
    \includegraphics[width = 0.49\textwidth]{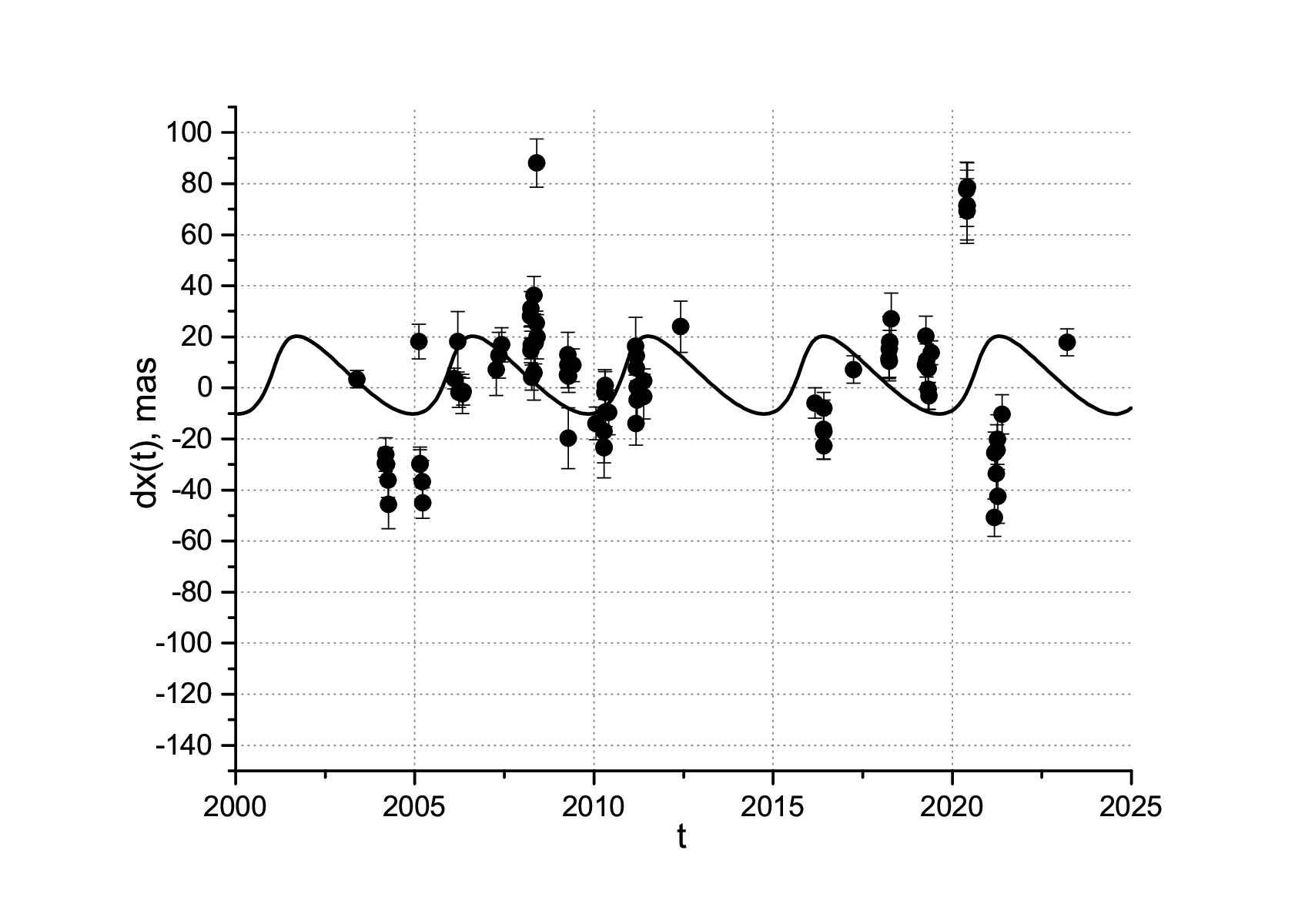}
    \includegraphics[width = 0.49\textwidth]{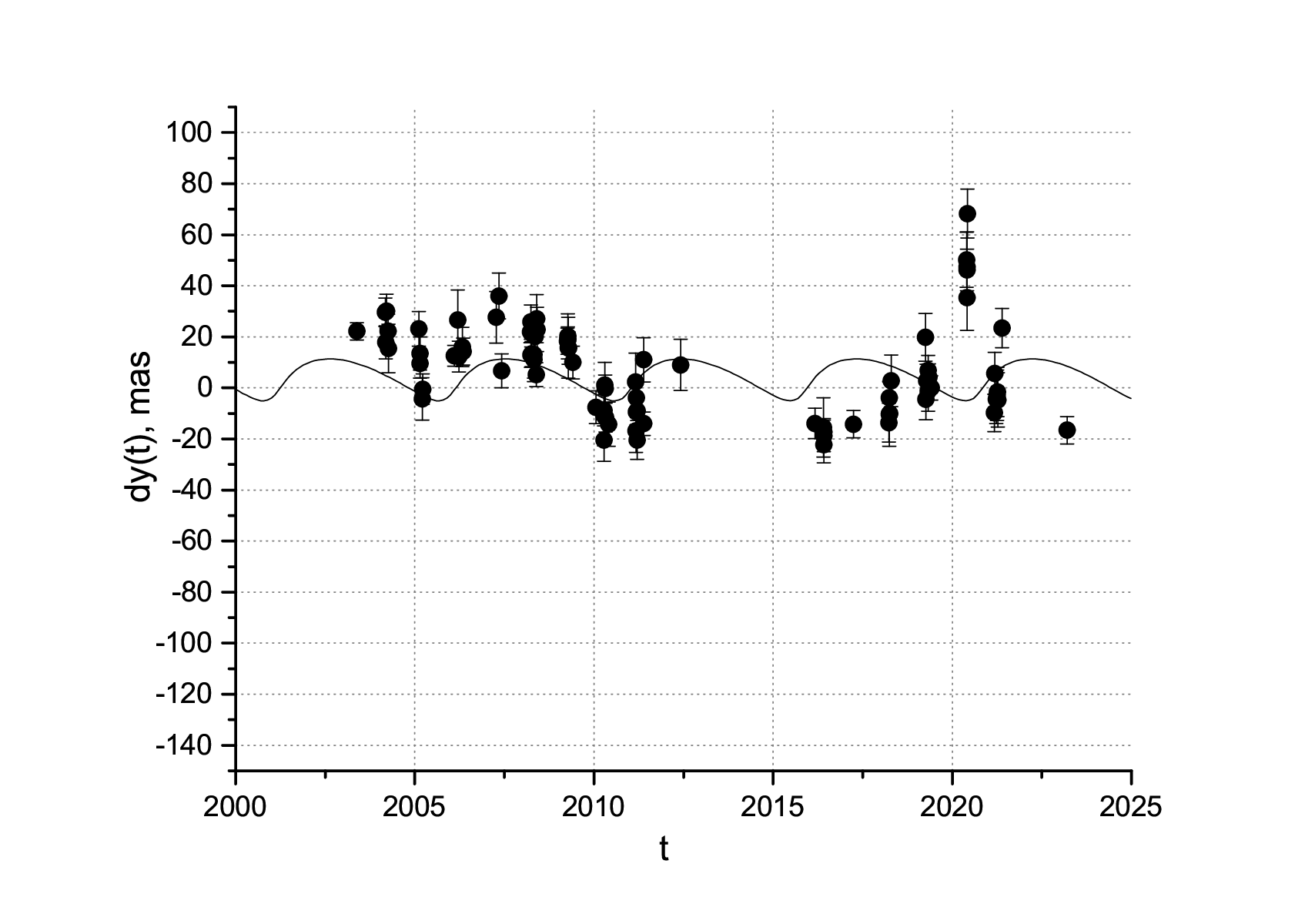}
\caption{Comparison of the Ba-Bb astrometric orbit ephemeris with deviations of individual CCD observations relative to the AB orbit.} 
\label{fig-dxy}
\end{figure}

The residuals are averaged over the phase in a sliding window equal to $0.05P\approx0.25$~years, the errors of each position are external and determine the convergence within the phase. Ephemeris of the inner pair Ba-Bb orbit in comparison with all initial data are also presented in Fig. ~\ref{fig-dxy}. 

The periastron longitude from the ascending node $\omega$ is determined independently from the spectroscopic and from the astrometric orbit, and therefore this parameter can be considered as a consistency control orbits. Within the error, the values of $\omega$ coincide. Thus, we have supplemented the spectroscopic orbit with the missing parameters $i$ and $\Omega$. 

\section{Is there another satellite in this system?}

Let us consider the residuals of Pulkovo CCD observations. After taking into account the orbit of the outer AB pair mean square deviations $\sigma_x=42.8$~mas, $\sigma_y=42.5$~mas. After taking into account astrometric orbit of Ba-Bb, they have significantly decreased ($\sigma_x=30.8$~mas, $\sigma_y=19.1$~mas). These residuals (dx2, dy2) are also presented in Table \ref{tab-dxy}. Here, the weights of observations are not taken into account. 

\begin{figure}
    \centering
    \includegraphics[width = 0.49\textwidth]{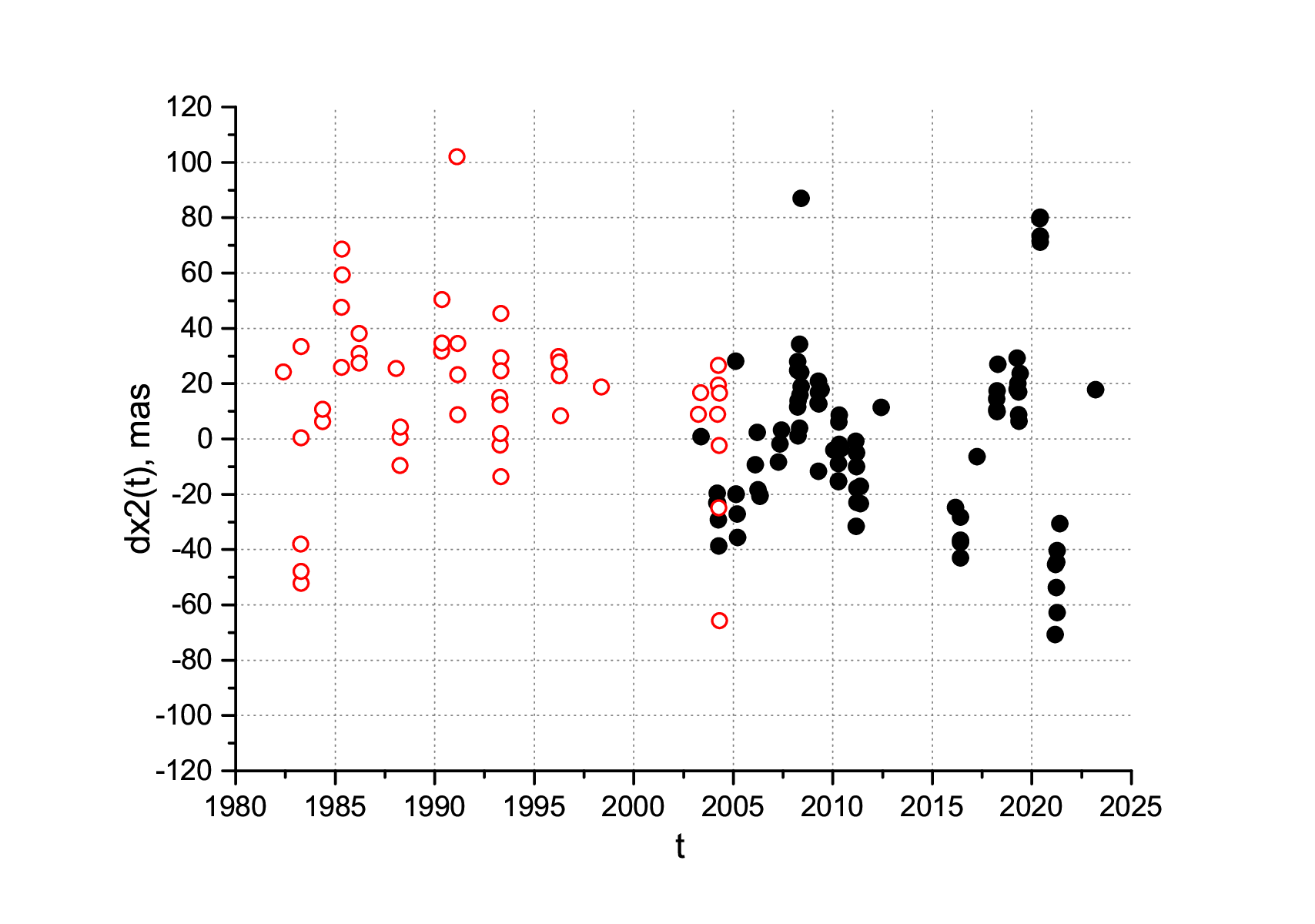}
    \includegraphics[width = 0.49\textwidth]{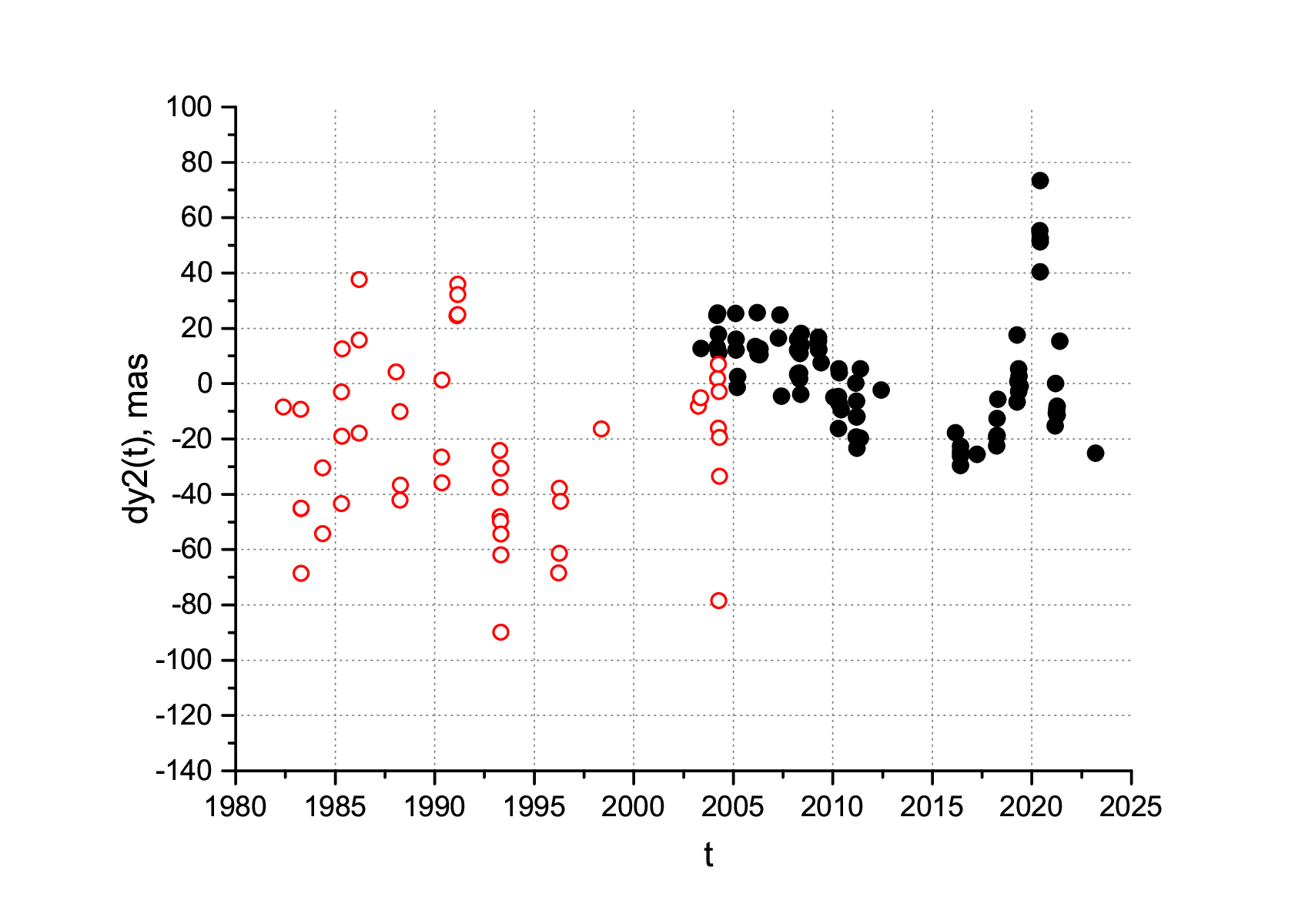}
\caption{Residual deviations of Pulkovo photographic (red open circles) and CCD (black circles) observations after taking into account both orbits: AB and Ba-Bb.}
\label{fig-dxy2}
\end{figure}

On Fig. \ref{fig-dxy2}, the deviations of Pulkovo photographic observations are also added to them after accounting for both orbits. It can be seen on the dy2(t) graph, that the move of photographic and CCD deviations is repeated in the time intervals 1985--1996 and 2005--2012. The characteristic trend allowed us to assume satellite presence with a period of approximately 20 years. Unfortunately, the star was rarely observed by photography after 1996, and 2012--2015 CCD observations are absent. Until 2019, CCD-observations do not contradict this assumption, but the latest observations are 2020, 2021 and 2023 made us doubt.

%\newpage
\begin{figure}
    \centering
    \includegraphics[width = 0.49\textwidth]{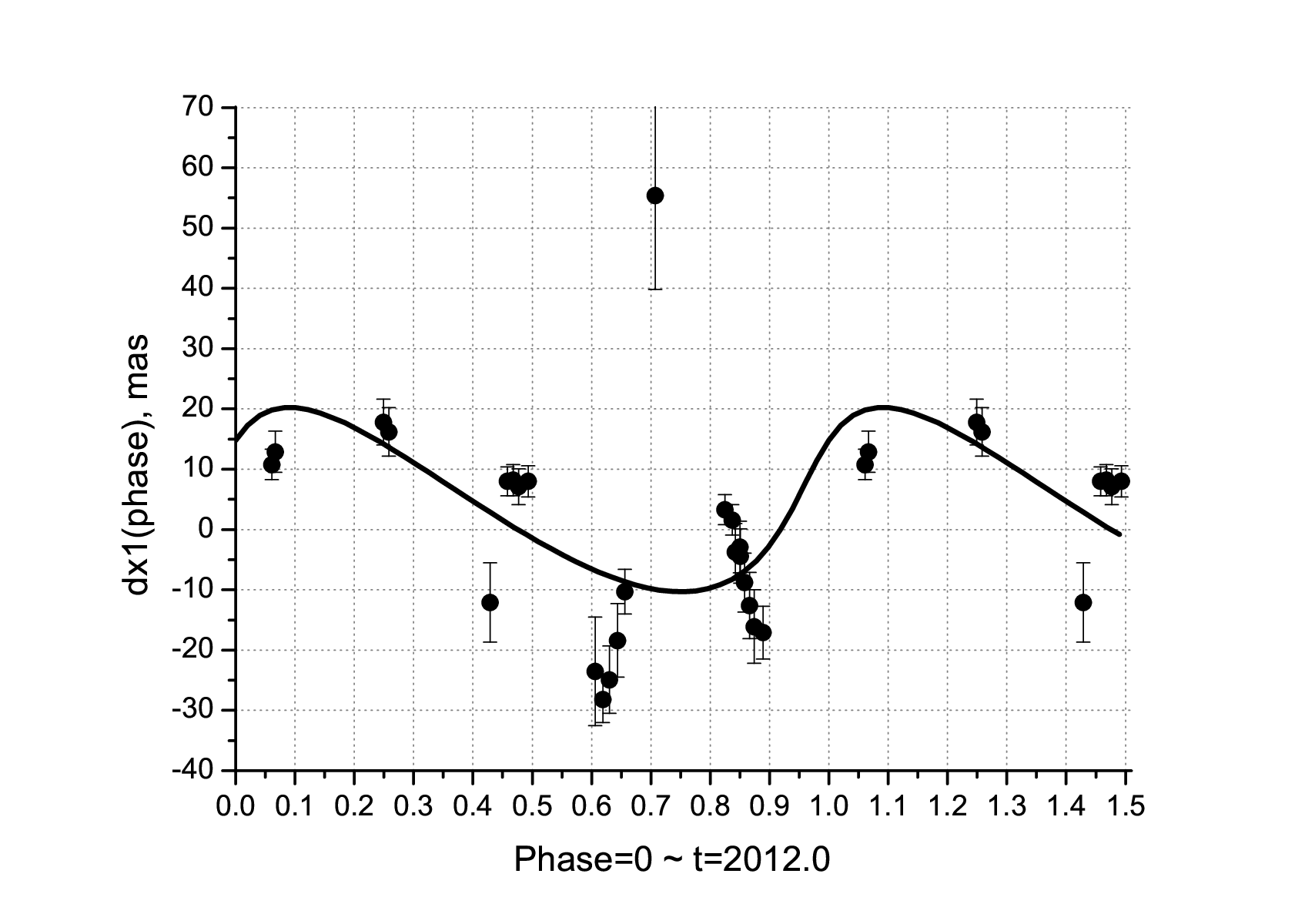}
    \includegraphics[width = 0.49\textwidth]{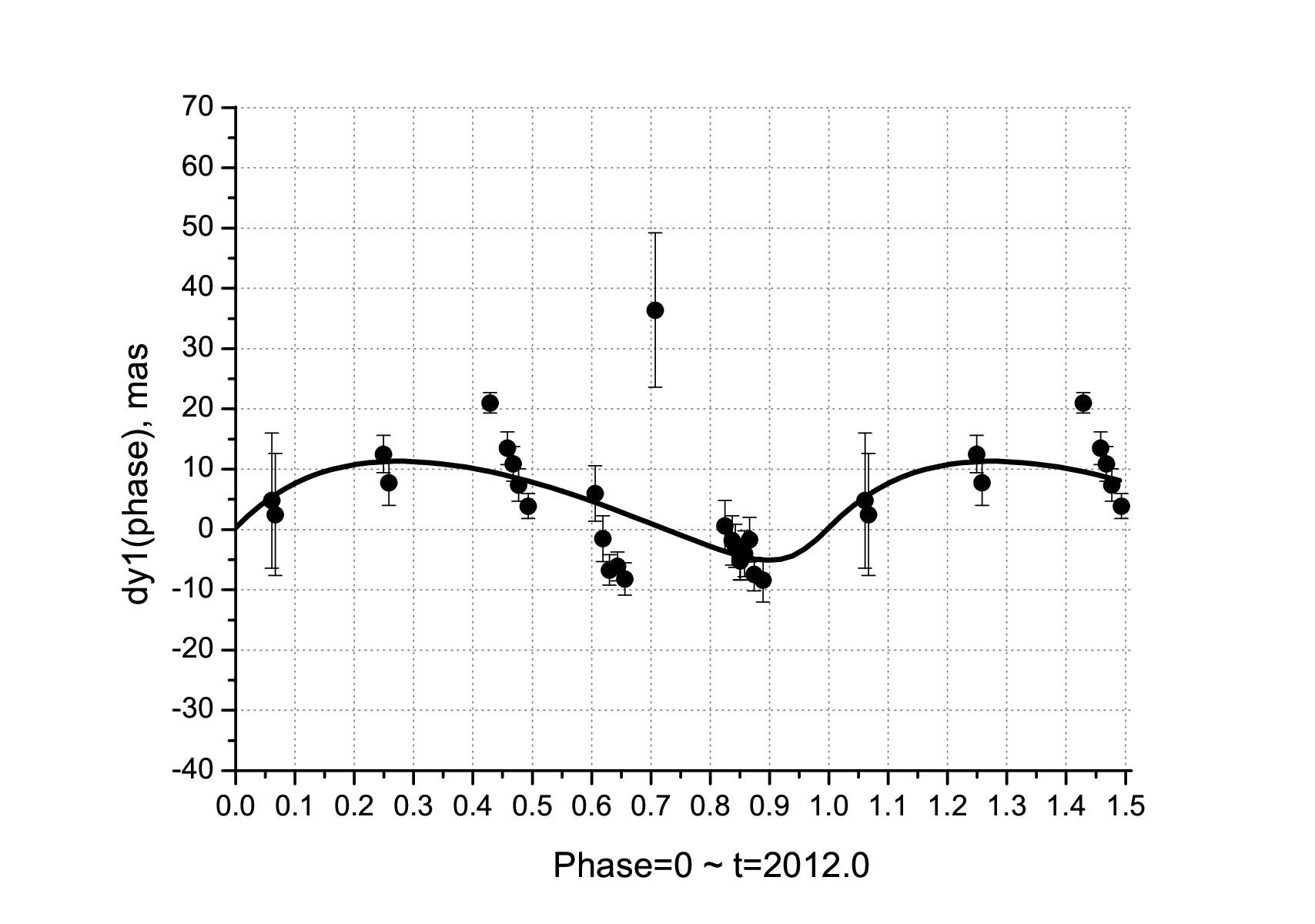}
\caption{Comparison of Ba-Bb astrometric orbit ephemeris with phase-averaged residuals of CCD observations relative to AB orbit.}
\label{fig-orbB}
\end{figure}

In addition, in Fig. \ref{fig-orbB}, we see that the positions near close phases differ from the natural random spread and indicate some systematic movement. 

Therefore, we have now come to the conclusion that the system has a short-period satellite (perhaps more than one) that we cannot confidently detect, and the 20-year period is caused by observational selection or is a multiple for several satellites. Compared to our Solar system, then an outside observer would most likely find a period of 60 years (the influence of Jupiter and Saturn). 

Thus, this question remains open for now. The CCD observations with the 26-inch refractor are continued. Also radial velocity observations and high-precision speckle-interferometric observations of the star ADS 9173 are needed.

\begin{table}
\begin{center}
\caption{Orbital elements of the Ba-Bb pair.} \label{tab-AB}
\begin{adjustbox}{width=0.99\textwidth}
\begin{tabular}{r rrrrrrrrccc}
\hline
     & $a_{ph}, mas$ & $a$, AU & $P$, yr & $e$ & $\omega, ^\circ$  & $i, ^\circ$ & $\Omega, ^\circ$ & $T$, yr &
 $M_{Ba}, M_\odot $ & $M_{Bb}, M_\odot $ & Reference \\
\hline
     &  18.3  & 3.56 & 4.904 &  .53 &  82 & 109 &  251 & 2010.917 & 1.4 & $\geq 0.48$ & This work \\
$\pm$ &  3.1  & - & - &   - &     15  &  14 &   10 &        -    &  -  &     -       &  \\
\hline
     &  *21.3 & - & 4.904 &  .53 &  96 & - &  - & 1927.583 & - & $\geq 0.5$ & \cite{Bakos} \\
$\pm$ &   3.3  & - & 0.009 &  .09 &   3 &  - &  - &  0.016 & - & - & \\
\hline
\end{tabular}
\end{adjustbox}
\end{center}
 {\scriptsize  Here: $a_{ph}$ is the semi-major axis of the photocenter orbit, $a$ is the semi-major axis of the relative orbit.  * according to \citep{Bakos} $a_{ph}*sin(i)=1$ au, we recalculated the $a_{ph}$ value taking into account the Gaia DR3 parallax \citep{G2022}, the inclination, and its error.}
\end{table}

\bibliography{references}
%\newpage

\end{document}